# The Peculiar Behavior of Halo Coronal Mass Ejections in Solar Cycle 24

N. Gopalswamy[1], H. Xie[1,2], S. Akiyama[1,2], P. Mäkelä[1,2], S. Yashiro[1,2], and G. Michalek[3]

[1]NASA Goddard Space Flight Center, Greenbelt, Maryland

[2]The Catholic University of America, Washington, DC

[3]Astronomical Observatory of the Jagiellonian University, Krakow, Poland

## ABSTRACT

We report on a remarkable finding that the halo coronal mass ejections (CMEs) in cycle 24 are more abundant than in cycle 23, although the sunspot number in cycle 24 has dropped by ~40%. We also find that the distribution of halo-CME source locations is different in cycle 24: the longitude distribution of halos is much flatter with the number of halos originating at central meridian distance ≥60$^o$ twice as large as that in cycle 23. On the other hand, the average speed and the associated soft X-ray flare size are the same in the two cycles, suggesting that the ambient medium into which the CMEs are ejected is significantly different. We suggest that both the higher abundance and larger central meridian longitudes of halo CMEs can be explained as a consequence of the diminished total pressure in the heliosphere in cycle 24 (Gopalswamy et al. 2014). The reduced total pressure allows CMEs expand more than usual making them appear as halos.

## 1. Introduction

A halo coronal mass ejection (CME) appears to surround the occulting disk of the observing coronagraph in sky-plane projection. Halo CMEs were first reported by Howard et al. (1982) and only a handful were recorded by the Solwind coronagraph onboard the P78-1 mission (Howard et al. 1985). Halo CMEs constitute only ~3%) of all CMEs and represent an energetic population because most of the CMEs that produce large solar energetic particle (SEP) events and major geomagnetic storms are halos (Gopalswamy et al. 2010a). Halo CMEs generally originate from close to the disk center, while ~10% originate close to the limb (Gopalswamy et al. 2010b). In limb halos, the disturbance appearing over the opposite limb is likely to be a shock (Gopalswamy et al. 2010c). Disk-center halos are responsible for severe geomagnetic storms and their geoeffectiveness declines as the central meridian distance (CMD) increases (Gopalswamy, Akiyama, and Yashiro, 2007). During cycle 23, the average sky-plane speed of halos was ~1100 km/s, more than two times that of regular CMEs. The true speed of halo CMEs is likely to be larger because the sky-plane speeds are subject to large projection effects (Gopalswamy et al. 2010b). The first hint that there is something different about cycle-24 halos was that all large SEP events of this cycle were associated with halos (Gopalswamy 2012; Gopalswamy et al. 2014a,b). Among limb CMEs associated with soft X-ray flares, cycle 24 had roughly three times more halos than in cycle 23. These differences motivated us to compare the properties of halo CMEs in cycles 23 and 24.

## 2. Observations

We consider all halo CMEs observed in cycle 24 (December 2008 to December 2014) and compare them with the cycle-23 halos observed during the corresponding 73 months: May 1996 to June 2002. These CMEs are all listed in the online halo CME catalog (http://cdaw.gsfc.nasa.gov CME_list/halo/halo.html, Gopalswamy et al. 2010b). Each halo is identified by its first-appearance time in the field of view (FOV) of the Large Angle and Spectrometric Coronagraph (LASCO, Brueckner et al. 1995) on board the Solar and Heliospheric Observatory (SOHO) mission. The first-appearance refers to the C2 coronagraph FOV (2.5 to 6 Rs). However, the halo designation is based on the appearance in the C3 coronagraph FOV (3 to 32 Rs) because some CMEs appear as partial halos in the C2 FOV and become full halos only in the C3 FOV. Also, we consider only the outermost disturbance in deciding whether a CME is halo or not, although both CME flux ropes and the surrounding disturbance can appear as halos (Michalek et al. 2007). The sky-plane speeds and accelerations are obtained at a position angle at which the CME appears to move the fastest (the measurement position angle, MPA). The solar sources of the CMEs (heliographic coordinates of the eruption region) have been identified as the location of the associated flare, filament eruption, or coronal dimming from various observations (see Gopalswamy, Akiyama, and Yashiro 2007 for details). Javascript movies combining LASCO images with GOES X-ray light curves are also used to confirm source locations. The space speeds are estimated by inputting the sky-plane speed and source coordinates in a cone model (Xie, Ofman, and Lawrence 2004). Finally, the X-ray importance and start time of the associated flare (if available) are given. It must be pointed out that the halo CMEs are subject to a selection effect in that only CMEs expanding rapidly to surround the occulting disk are identified as halos. For a given coronagraph (LASCO C3 here), halos represent an energetic population. We are mainly concerned with the occurrence rate, source locations, and speeds of the halo CMEs in cycles 23 and 24.

### 3. Analysis and Results

In this section, we consider the occurrence rate, CME sources (longitudes and latitudes), and CME speeds. The width of the halo CMEs are unknown, so the speed serves as the main indicator of how energetic a CME is. We also the space speeds from the cone model (Xie, Ofman and Lawrence 2004) for comparison. Using STEREO and SOHO data for a set of 20 Earth-directed CMEs reported in Gopalswamy et al. (2013), we show that the space speed of CMEs estimated from the cone model is consistent with space speed directly measured from STEREO images.

### 3.1 Halo CME Rate

Figure 1 shows the number of halos per Carrington rotation observed by SOHO/LASCO. There were **260** halos in cycle 24, with a similar number (221) over the corresponding phase in cycle 23. Excluding the 5-month gap in LASCO observations (24 June 1998 to 15 October 1998 and 20 December 1998 to 5 February 1999), we obtain the monthly rate of halo CMEs as 3.25 based on the 68 months of SOHO observation in cycle 23. Assuming that halos occurred with the same rate during the data gap, we estimate a total number of 237 halos in the first 73 months of cycle 23. There were additional 175 halos during the rest of cycle 23, corresponding to a rate of ~2.2 halos per month. While we have accounted for extended interruption in SOHO operations, there were also data gaps that might affect the halo count. Based on the extent (2.5 to 32 Rs) of the LASCO FOV and the average halo speeds (~1000 km/s), we estimated that ≥6-hour data gaps are likely to affect the halo count. We corrected for such data gaps as follows. If N is the observed number of halos, $T_d$ the down time and $T_u$ the up time in a 3-month interval, then the corrected number $N_c = N + N*T_d/T_u$ in the interval. For the three-month period without observations, an average of the preceding and succeeding three-months was used. After the correction, the total number of halos remained similar in the two cycles: 264 (cycle 23) and 267 (cycle 24). Occasionally, it is possible that many simultaneous eruptions from different position angles may give the impression of a single halo. However, the chance is extremely low because the EUV images available simultaneously can tell the locations of smaller CMEs (unless they are far behind the limb). Also during periods of high activity, it is possible that a few faint halos may be missed, but this applies to both cycles.

It is significant that the halo CME rate in cycle 24 has not decreased compared to that in cycle 23 even though the sunspot number (SSN) declined by ~40%: 75.98 in cycle 23 vs. 45.69 in cycle 24, averaged over 73 months. The normalized halo CME rates are 0.080/SSN (cycle 24) and 0.048/SSN (cycle 23). In other words, the halo CME occurrence rate did not drop in cycle 24 unlike SSN. The increase in LASCO image cadence in August 2010 did not affect the halo count. When we reduced the cadence by a factor of 2, none of the 41 halos in 2011were missed and were detected in 4 to 25 frames depending on the CME speed. The result that the daily CME rate tracked the sunspot number when corrected for the cadence change (Wang and Colaninno 2013) certainly does not apply to halos. During the early phase of the SOHO mission (before 1999) the image cadence was low and irregular; the FOV was also truncated. We estimated that these do not affect the halo count based on the average speed of halo CMEs and the time they take to cross the LASCO FOV.

We also note that the halo CME rate has the largest peak coinciding with the second peak in SSN. The interval between the two peaks in solar activity maxima is known as the Gnevyshev gap (Gnevyshev, 1967), although the gap is not universal in all phenomena showing solar cycle variation (Kane, 2006). The second SSN peak is more pronounced in cycle 24, which seems to be generally rare (only 5 such cycles in the past 215 years: 5, 7, 12, 16, 24.) Interestingly, all the cycles with dominant second peak are small cycles (SSN peak around 50). Cycle 24 closely resembles cycle 12 in this behavior.

### 3.2 Halo CME Source Locations

There was a slightly larger fraction of frontside halos in cycle 23 (147 out of 221 or 66.5%) than that (128 out of 260 49.2%) in cycle 24. Part of this is likely to be due to statistical chance. In a few percent of cases, the simultaneous occurrence of a backside halo and a frontside eruption with no obvious CME in the LASCO FOV would be classified as a frontside halo. This problem is mitigated in cycle 24 because of the availability of STEREO views from off the Sun-Earth line.

Figure 2 (a,e) shows that the halos are generally confined to the active region belt because actives regions possess the energy required to power these energetic CMEs (Gopalswamy et al. 2010b). The cycle-24 halo sources are more uniformly distributed in longitude than those in cycle 23. Source location close to the central meridian makes the CME more likely appear as a halo and only occasionally CMEs from the limb become halos if they are very energetic and the associated shocks are observed above the opposite limb. The longitudinal distribution of halo CMEs sources in cycle 24 is different from that in cycle 23.

The latitude-time plots Fig. 2 (b,f) also show some differences between the two cycles. The cycle-23 latitudes did not have north-south asymmetry in the beginning of the cycle, while in cycle 24, the halos originated mostly in the northern hemisphere. The maximum phase was spread over three years in both cycles (1999-2002.5 in cycle 23 and 2011 to 2014.5 in cycle 24). The activity peaked first in the north and then in the south in both cycles, the lower activity between the peaks corresponding to the Gnevyshev gap in SSN. The Gnevyshev gap is wider in cycle 24, because one can see the concentration of halo CME sources in the northern latitudes towards the end of 2011 and around the turn of 2014 in the southern latitudes. The gap is shorter in cycle 23 with the concentration of halo CME sources around the turn of 2001 in the northern latitudes and in early 2002 in the southern latitudes.

The longitude and latitude distributions in Figure 2(c,d) further confirm the flatter longitudinal distribution in cycle 24. Out of the 128 cycle-24 front-side halos, 40 (or 31%) were at CMD ≥60°, compared to only 24 out of the 147 halos (or 16%) in cycle 23 (see Fig. 2f). The limb halos in cycle 24 are two times more abundant than those in cycle 23. We performed the student's t-test on the absolute values (i.e., CMD) of the halo CME longitudes in the two cycles. The range of means in the two cycles did not overlap (95% confidence intervals were 24°.88 through 33°.41 for cycle 23 and 33°.55 through 42°.67 for cycle 24), confirming that the longitude distributions are significantly different (Fig. 2(c,g)). The probability P that the two distributions are different by chance is 0.005. The difference was also confirmed to be significantly different using the Kolmogorov-Smirnov test (P = 0.006). The latitude distributions are also significantly different between the two cycles: the cycle-23 distribution was significantly wider as confirmed by both student's t-test and Kolmogorov-Smirnov test (P= 0.0056 and 0.005, respectively). The 95% confidence intervals of the mean did not overlap in both the tests. The difference in the latitude distribution is due to the appearance of the active region belts in the two cycles. However, the difference in the longitude distribution is puzzling, which we shall discuss further in the next section.

### 3.3 Halo CME speeds

Halo CMEs are known to be faster than the average CMEs by a factor of at least 2 and are generally associated with larger flares (Gopalswamy, Yashiro, and Akiyama, 2007; Gopalswamy et al. 2010b). For a set of 314 cycle-23 halos, the mean speed and flare size were ~1109 km/s and M1.0, respectively (Gopalswamy et al. 2010b). The average speed from the cone model increased to 1296 km/s.

There was no independent verification of the space speeds reported earlier from the cone model (Gopalswamy et al., 2010b). Fortunately, SOHO and STEREO were in quadrature during 2010-2012, at which time many front-side halos were observed by SOHO/LASCO. Gopalswamy et al. (2013) selected 20 LASCO halos such that their solar sources were within 30° from the limb in the view of one of the STEREO spacecraft. It was possible to obtain the space speed of these CMEs using STEREO/COR2 observations, since the projection effects were minimal. These speeds were higher than the LASCO speeds, which are subject to projection effects. We also estimated the space speeds from the cone model with LASCO speeds and source locations as input. Figure 3 compares the sky-plane speed from LASCO with the two space speeds (from STEREO observations and cone model). The cone model speeds were always greater than the LASCO speeds. This was also true for the STEREO speeds, except for a couple of events. The correlation coefficients (r) were high for both sets: r =0.98 for the cone model speed with LASCO speed and r = 0.87 for the STEREO space speed with LASCO speed. The two space speeds were also correlated (r = 0.80). This analysis confirms that the cone model speed is a reasonable approximation to the space speed of halos.

Figure 4 shows the CME speed and flare size distributions are very similar between the two cycles: the difference in average speeds is much less than the typical errors resulting from height-time measurements (~10%). The mean and median flare sizes are the same for the two cycles. The speeds are also similar to the ones obtained for the whole of cycle 23 (Gopalswamy et al. 2010b). Thus we conclude that there is nothing unusual about the halos in cycle 24 as far as the speed and flare size are concerned.

Figure 5 shows the variation of the sky-plane speed with source location (CMD). The speed increased with CMD primarily because of projection effects. This is clear from the cone-model speeds over-plotted (open circles). But there is also some selection effect involved because the limb halos have to be more energetic for the disturbance to show up above the opposite limb. The difference between the sky-plane speed and space speed is larger for smaller CMD (close to disk center) and the difference almost disappears after CMD = 60°. But what is clear in the plots is the larger number of limb halos in cycle 24, which was also noted in Fig. 2.

**4. What is Different in Solar Cycle 24?**

The opposite behavior of SSN and halo CME rate is a puzzling result, especially because halo CMEs mostly originate from sunspot regions (a small fraction of halos do originate in quiescent filament regions). On the other hand, CME kinematics and the flare-size distributions do not differ significantly between cycles 23 and 24. This suggests that the difference may be due to the ambient medium into which the CMEs are ejected. While the weak solar activity did not affect the halo CME rate, it did affect the state of the heliosphere. Gopalswamy et al. (2014) reported that the total pressure in the heliosphere decreased by about 40% and they suggested that the anomalous expansion of cycle-24 CMEs can be explained by the reduced total pressure. They found that for a given speed, the cycle-24 CMEs were wider than those in cycle 23. For a given internal pressure of a CME flux rope, a decrease in external pressure would result in a larger size of the CME, which explains the larger size of the cycle-24 CMEs. This expansion can readily account for the increased number of halos at larger CMD and hence the overall increase in the number of halo CMEs. Rapid expansion of the CMEs is one of the key processes that determine whether a CME is a halo or not for a given coronagraph. The expansion caused by the reduced ambient pressure is consistent with the increase in the number of halos overall and the increase in the number of halos originating at larger CMDs. A related issue is the reduced Alfven speed in the corona due to the reduced magnetic field strength and the density, makes it easier to form shocks (Gopalswamy et al. 2014). Since we use the outermost structure (including the shock) in deciding whether a CME is a halo or not (Yashiro et al., 2008), the easier shock formation might have also contributed to the higher abundance of halos in cycle 24.

## 5. Summary and Conclusions

Using SOHO/LASCO data covering solar cycles 23 and 24, we compared the properties of halo CMEs in the two cycles. The study period corresponds to the first half of the cycles, up to the end of the maximum phase (the solar activity peaked in the beginning of 2014 and is already in the declining phase). The sunspot number averaged over the 73 months dropped by ~40% in cycle 24 relative to cycle 23. However, the number of halo CMEs did not decrease in cycle 24. There were 221 halo CMEs in cycle 23 and 260 in cycle 24. Taking into account of the SOHO disability periods in cycle 23 and f ≥6-hour data gaps in the two cycles, we estimated the corrected halo counts to be 264 (cycle 23) and 267 (cycle 24). Thus the halo CME activity did not decrease in cycle 24 in contrast to the sunspot number, which dropped significantly. In fact, if we normalize the halo CME rate to the SSN, then the cycle-24 rate per SSN is about 67% higher.

The main conclusions of this study are:

(i) Halo CMEs are as abundant in cycle 24 as in cycle 23, even though the sunspot number dropped significantly in cycle 24.

(ii) The source distribution of halo CMEs on the disk are significantly different in cycle 24: the number of halos originating from CMD≥60$^{o}$ is twice as large as that in cycle 23.

(iii) The CME kinematics and the flare-size distribution of halos do not differ significantly between cycles 23 and 24. This suggests that the difference may be due to the ambient medium into which the CMEs are ejected. Different ambient properties in the two cycles seem to have modified the CME widths.

(iv) Both the higher abundance and larger average CMD of cycle-24 halos can be attributed to the weak state of the heliosphere: the reduced total pressure (plasma + magnetic) near the Sun allows CMEs expand more, making them appear wider in cycle 24 for a given speed.

**Acknowledgments** SOHO is a project of international collaboration between ESA and NASA. STEREO is a mission in NASA's Solar Terrestrial Probes program. The work of NG, SY, SA was supported by NASA/LWS program. PM was partially supported by NSF grant AGS-1358274 and NASA grant NNX15AB77G. HX was partially supported by NASA grant NNX15AB70G. GM was supported by NCN through the grant UMO-2013/09/B/ST9/00034. The authors thank the anonymous referee for helpful comments.

**Figures**

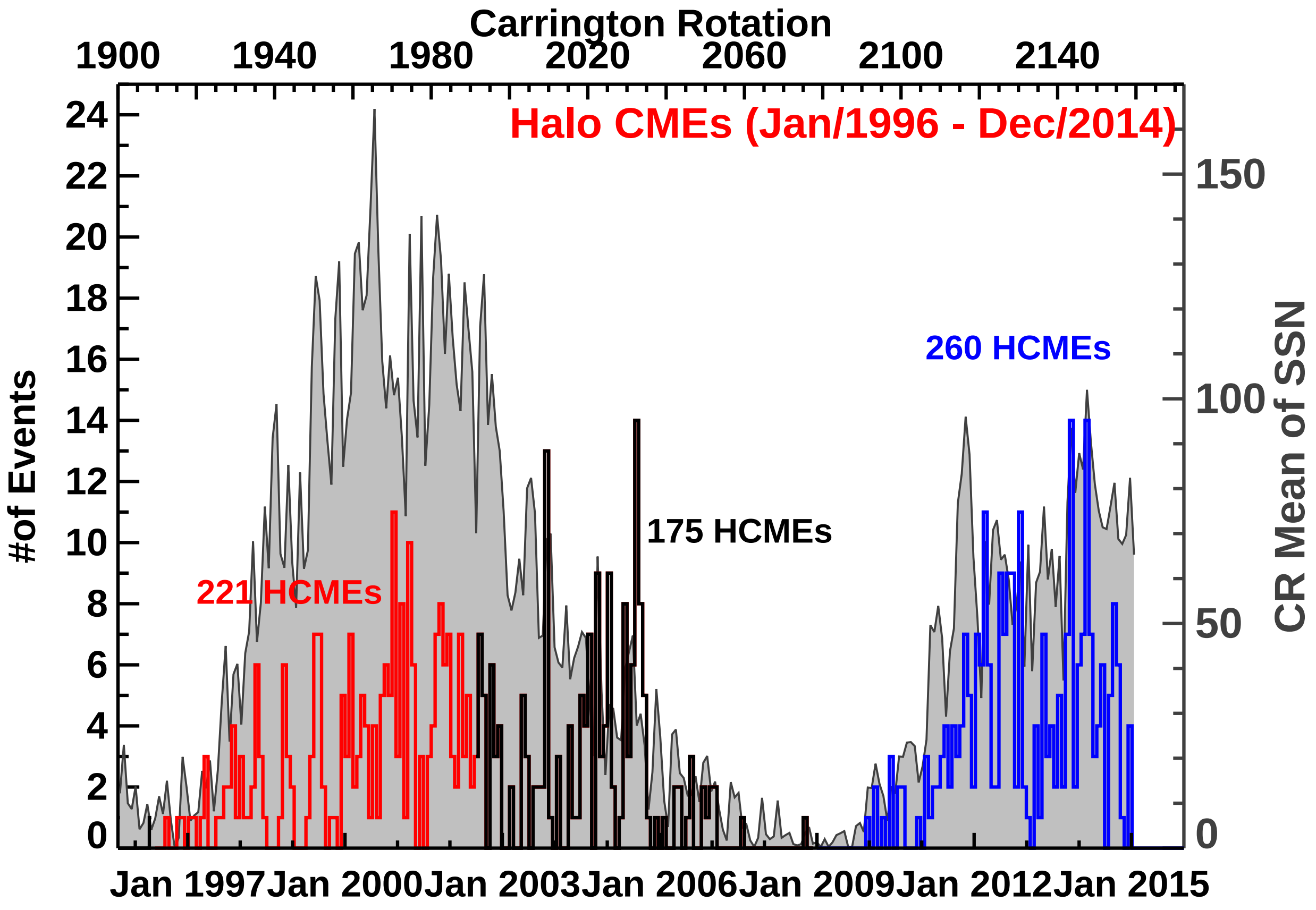

Figure 1. The number of halo CMEs (HCMEs) as a function of time binned over Carrington rotations (CRs). The rotation numbers are given at the top of the plot. The cycle-24 halos and the ones in cycle 23 over the corresponding epoch (first 73 months) are distinguished by the blue and red colors, respectively. The black histogram shows the halo CMEs outside the study period (one halo CME in cycle 22 (in the beginning of 1996) and 175 halos after May 2002 until the end of cycle 23). The number of halos in each of the intervals is shown on the plot.

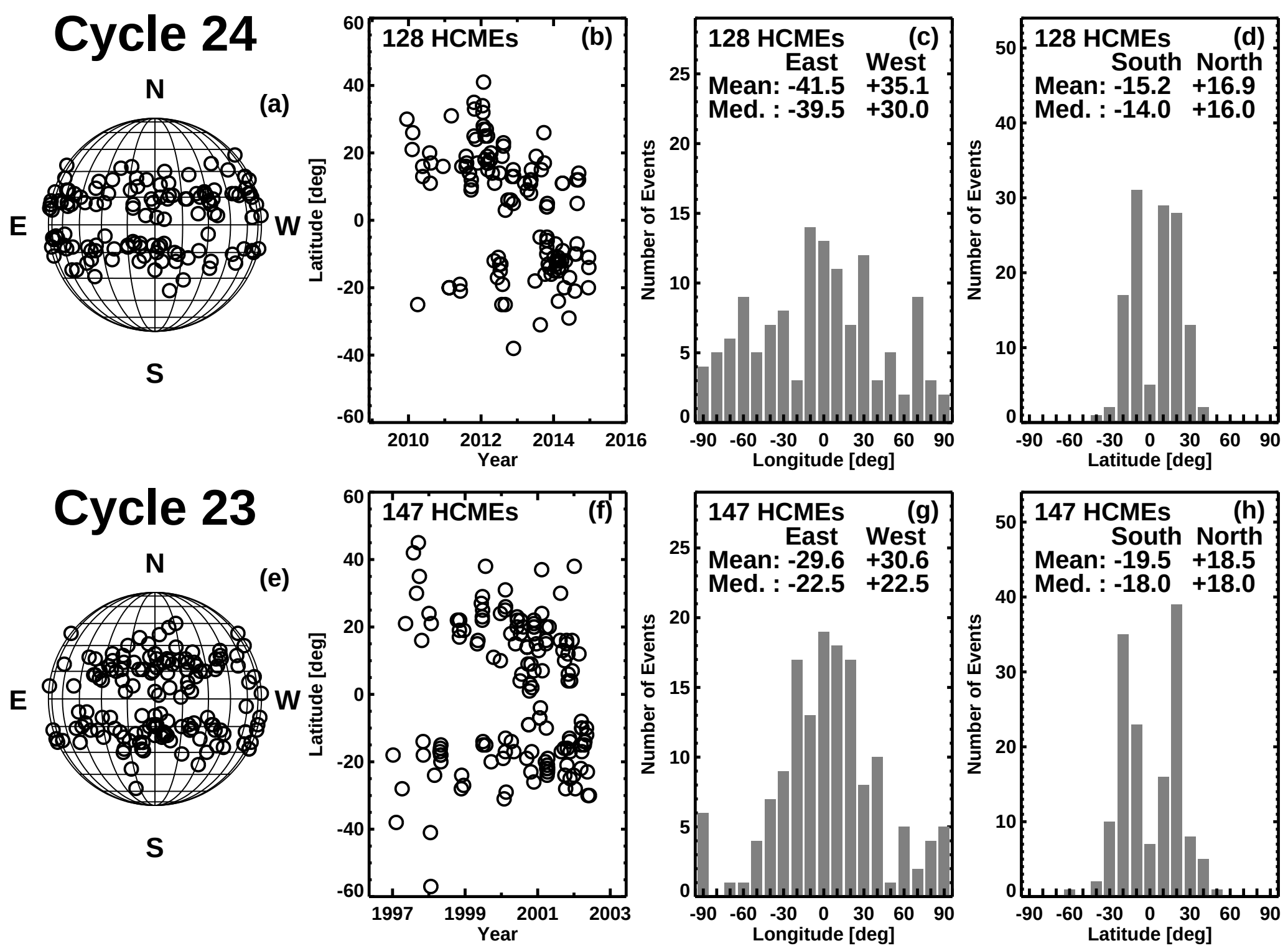

Figure 2. The heliographic coordinates (a,e) and the solar-cycle variation of source latitudes (b,f), source longitudes (c,g), and source latitudes (d,h) of halo CMEs in cycles 24 (top row) and 23 (bottom row). The mean and median values are shown on the plots separately for each hemisphere (eastern and western for longitudes; southern and northern for latitudes).

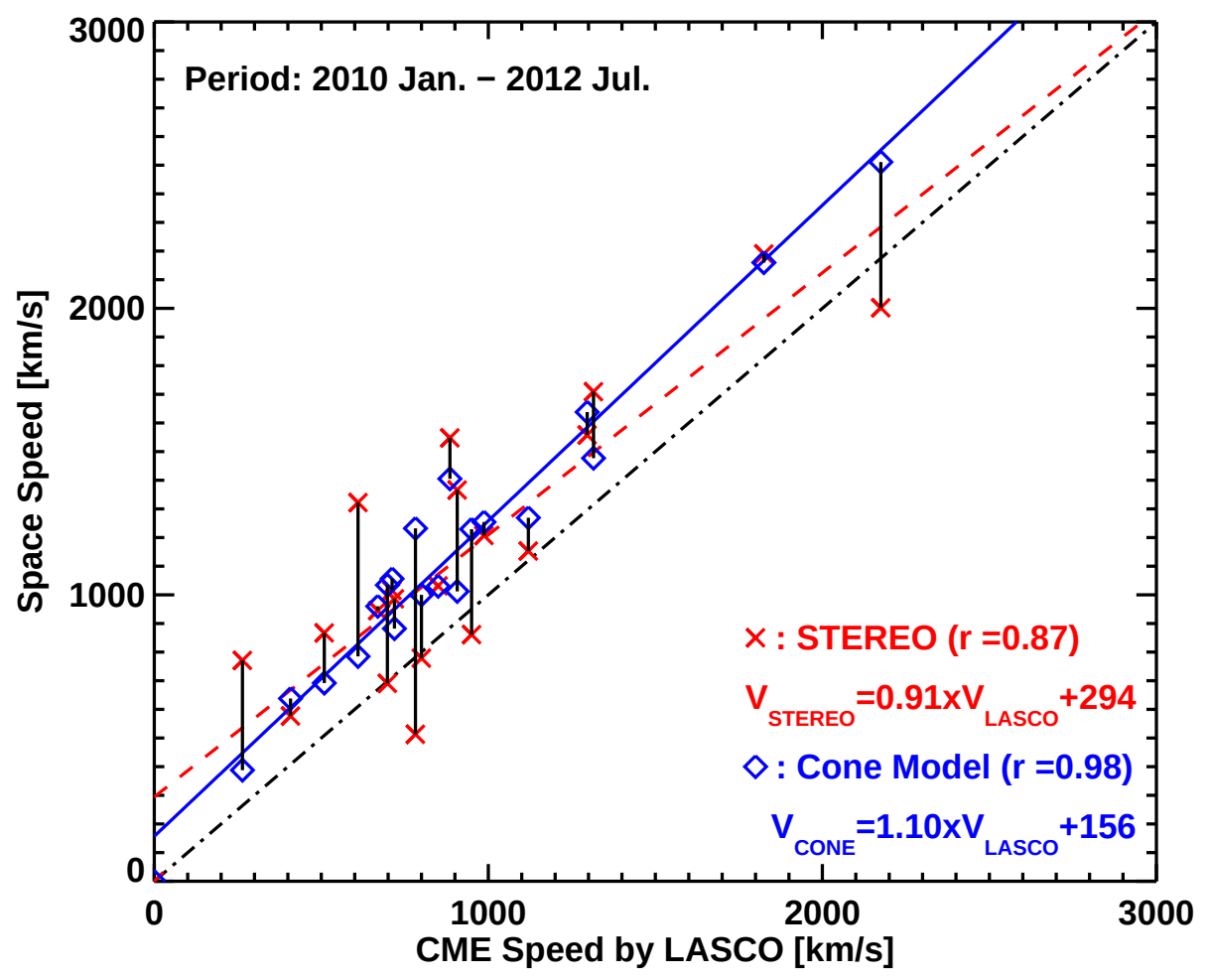

Figure 3. Scatter plot between the space speed and the LASCO sky-plane speed (crosses: measured space speed from STEREO; diamonds: cone model speed). The difference between the two space speeds is indicated by the vertical lines connecting the data points. The regression lines and the correlation coefficients are shown on the plot. The dot-dashed line is the bisector.

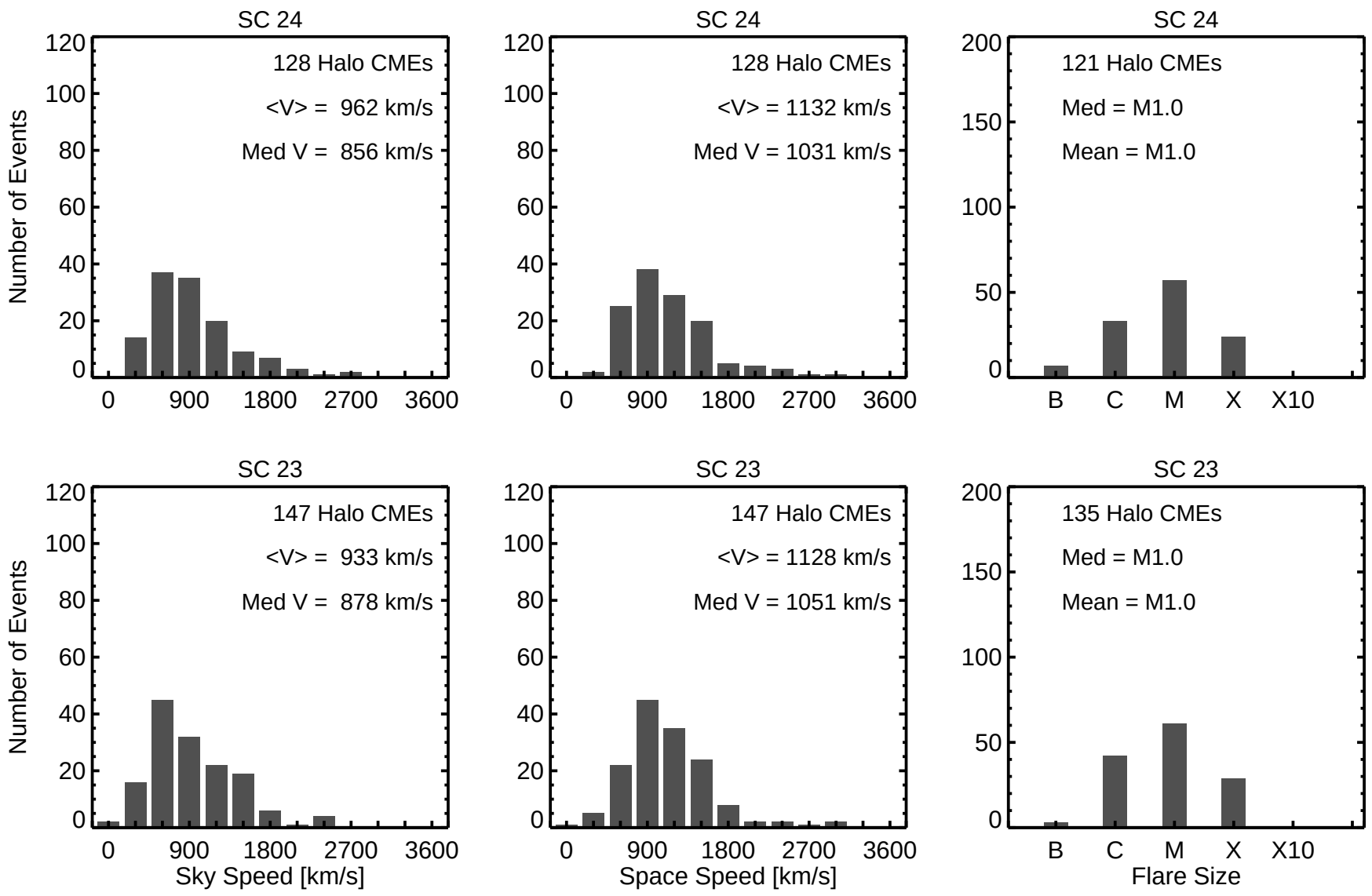

Figure 4. Comparison of speed and flare-size distributions of halo CMEs in cycles 23 and 24. (left) sky-plane speeds, (middle) space speeds from the cone model, (right) distributions of flare sizes.

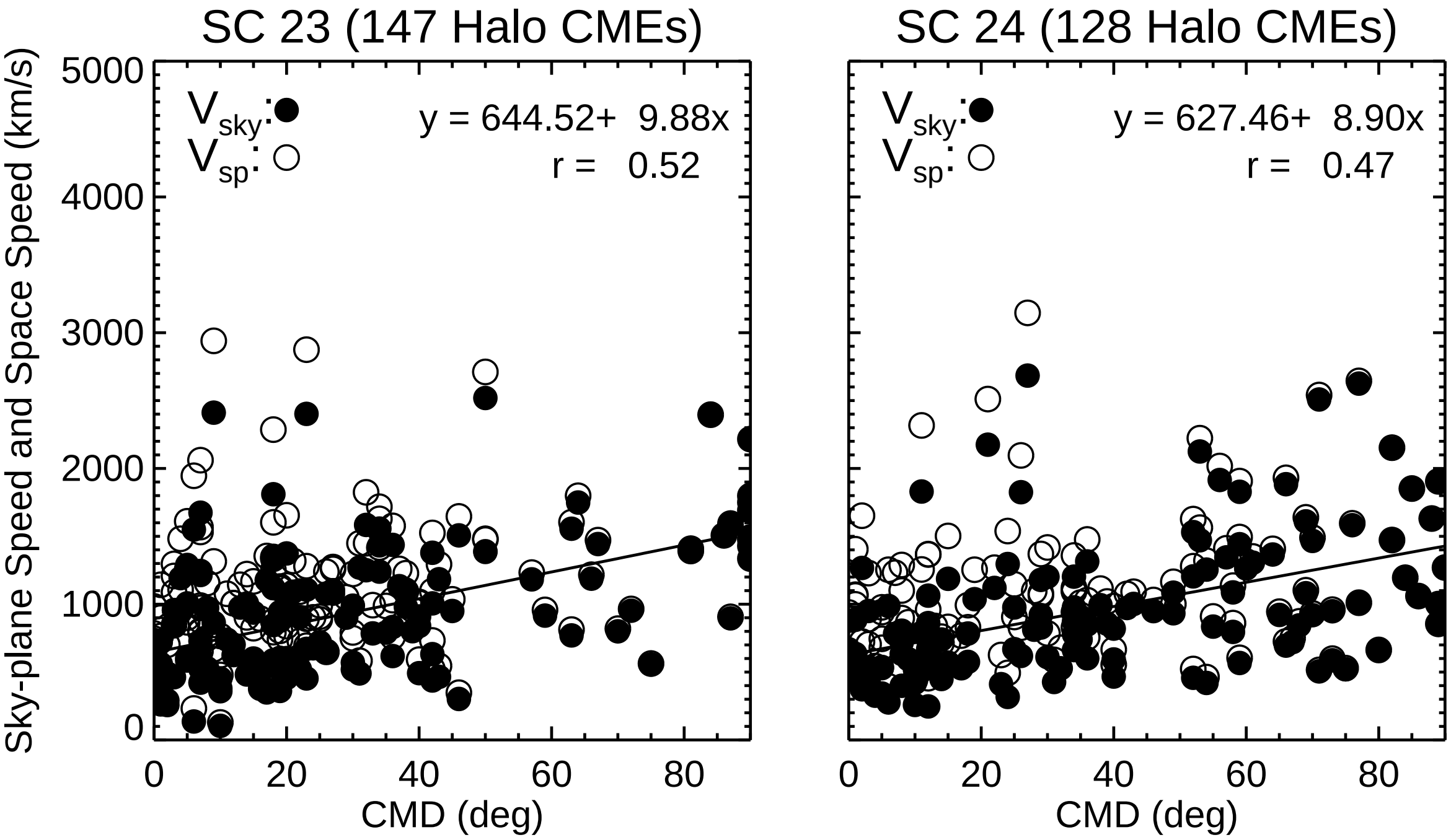

Figure 5. Scatterplot between CME speeds and the central meridian distance (CMD) for the cycle 23 and 24 halo CMEs. Filled and open circles represent sky-plane ($V_{sky}$) and cone model ($V_{sp}$) speeds, respectively. The correlation coefficients (CMD vs.$V_{sky}$) and the regression lines are very similar in the two cycles.